\documentclass[%
 reprint,
 amsmath,amssymb,
 aps,
]{revtex4-1}

\usepackage{graphicx}
\begin{document}


\title{Live Young, Die Later: Senescence in the Penna Model of Aging}
\author{Avikar Periwal}
\affiliation{Montgomery Blair High School, 51 University Blvd East, Silver Spring Maryland 20901}
\email[]{avperiwa@mbhs.edu}

\date{\today}

\begin{abstract}
Cellular senescence is thought to play a major role in age-related diseases, which cause nearly 67\% of all human deaths worldwide. Recent research in mice showed that exercising mice had higher levels of telomerase, an enzyme that helps maintain telomere length, than non-exercising mice. A commonly used model for biological aging was proposed by Penna. I propose two modifications of the Penna model that incorporate senescence and find analytical steady state solutions following Coe, Mao and Cates. I find that models corresponding to delayed senescence have younger populations that live longer.
\end{abstract}


\maketitle

\section{Introduction}
From early alchemists looking for the elixir of life, to modern day researchers, humans have always wanted to understand aging.  As people age, their cells go through replication cycles.  Each replication reduces the length of the telomeres in the cells.  If a cell's telomeres are too short, it may not be able to replicate\cite{herbig}.  Cells that can no longer replicate are termed senescent.  Cellular senescence is thought to play a major role in age-related diseases, which cause nearly 67\% of all human deaths worldwide\cite{degrey}.  Recent research in mice showed that exercising mice had higher levels of telomerase, an enzyme that helps maintain telomere length, than non-exercising mice\cite{werner}.  In humans, runners had longer telomeres than non-runners\cite{werner}.  Since shortened telomeres are thought to be related to death, this research would seems to indicate that people who exercise live longer lives.  Population studies do indeed show this, however, most studies show that only the 
mean lifespan increases in exercising populations, not maximum lifespan\cite{holloszy,sarna}. 
\par
A commonly used model for biological aging was proposed by Penna in 1995\cite{penna, stauffer}.  The model looks at death from a mutation  accumulation standpoint.  In the last 10 years, papers have been published describing methods for finding the age distributions created by the Penna model without actually simulating the model\cite{coe1,coe2,coe3}.  However, the research on senescence indicates that age-related death is not caused by mutation accumulation, but rather by an inability to reproduce after a number of cycles
.  I propose two modifications of the Penna model that use this mechanism instead of a mutation accumulation, and show that a mean lifespan can increase without affecting the maximum lifespan, which cannot be done by changing parameters in the original Penna model.
\section{Calculations}
The Penna model assigns a bit-string to each individual in the population.  Each bit corresponds to a timestep of the simulation.  A 1 in the string represents a mutation, and a 0 means no mutation.  If an individual has gone through $T$ 1s, then it dies. Each individual can have offspring, with probability $b$.  The child's bit-string is derived from the the bit-string of the parent, where each 0 has probability \it m \rm of becoming a 1.  The Penna model ignores positive mutations, because they are rare.  The length of the bit-string provides a hard limit for lifespan.  In this paper, I will work with only the $T=1$ case.
\par
\begin{figure}
 \includegraphics[scale=0.35]{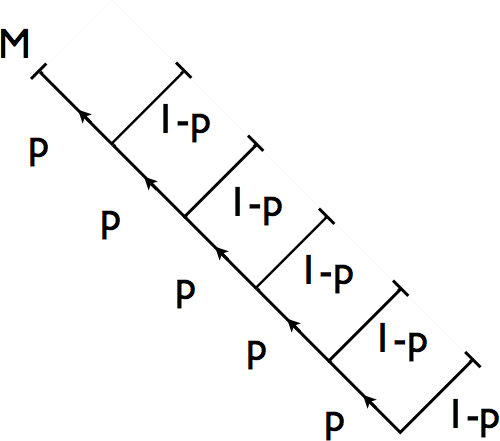}
 \caption{\label{treediagram} The possible paths until death, after reaching $l$, in the SP.}
\end{figure}
In the first proposed modification, which we will call the senescent Penna (SP) model, each individual can only get one disease, which is essentially the beginning of aging.  After the individual starts to age, it has probability $p$ of staying alive to reproduce at each time step.  The maximum number of replication cycles is $M$.  This is the Penna model, except instead of definitely dying, the individual has only a probability of dying.  People who exercise generally have longer telomeres, so once senescence starts, they have a smaller probability of dying.  Higher values of $p$ represent exercising populations.
\par
The first 1 in an individual's bit string is the age, $l,$ at which senescence begins.  The number of people alive at time  $j$, with age $x$, is $n_j(x,l,m)=pn_{j-1}(x-1,l,m-1)$ if $x>l$, where $p$ is the probability of living after senescence, and $m$ is the time since the inception of senescence.  If $x\leq l$, then $n_j(x,l)=n_{j-1}(x-1,l)$.  If $b$ is the probability of birth, and $e^{-\beta}$ is the probability of an individual going unmutated, then the number of children born in the next time step with disease acquisition time $l$, $n_{j+1}(0,l,0)$ is given by 
\begin{equation}
\begin{aligned}n_{j+1}(0,l,0)&=be^{-\beta l}\sum_x\sum_m n_j(x,l,m)\\&+(1-e^{-\beta})be^{-\beta l}\sum_{l'>l}\sum_x\sum_m n_j(x,l',m)\end{aligned}\end{equation}
Since $m$ is just the maximum of 0 and $x-l$, \begin{equation}\sum_x\sum_m n_j(x,l,m)=\sum_{x=0}^{l-1}n(x,l) + \sum_{x=l}^{M+l-1}n(x,l)\end{equation} where M is the maximum number of replication cycles allowed after disease.  Assuming a steady state, $n_{j+1}(x,l)=n_j(x,l)$.  Therefore, \begin{equation}\sum_{x=0}^{l-1}n(x,l,0)=l\cdot n(0,l,0),\end{equation} since each $x$ has the same number of people.  Since $n_j(x,l)=pn_{j-1}(x-1,l)$ for $x>l$, and since $n_{j+1}(x,l)=n_j(x,l)$ in a steady state \begin{equation}\sum_{x=l}^{M+l-1}n(x,l)=n(0,l)\sum_{x=1}^M p^x=p\frac{1-p^M}{1-p}n(0,l)\end{equation}  Defining $q_l$ as  \begin{equation}q_l=l+p\frac{1-p^M}{1-p},\end{equation}  and $n(l)=n(0,l)$, eq. 1 can be simplified to \begin{equation}0=be^{-\beta l} n(l)-\frac{n(l)}{q_l}+(1-e^{-\beta})be^{-\beta l}\sum_{l'>l+1}n(l')\end{equation}  Writing the same equation for $l+1$, some algebra leads to 
\begin{equation}n(l+1)=n(l)
\frac{be^{-\beta l}-1/q_l}
{be^{-\beta (l+1)}-e^{\beta}/q_{l+1}}
\end{equation}
\par
This equation leads to some limiting cases, in order to maintain a steady state. Neither the numerator, nor the denominator should vanish in eq. 7.\begin{equation}q_{max}<\frac{1}{1-e^{-\beta}},\end{equation} (and $l_{max}=q_{max}-p\frac{1-p^M}{1-p}$) and \begin{equation}b=\frac{1}{q_{max}e^{-\beta l}}\end{equation}
However, eq. 7 only gives the time of senescence, not the lifespan.  In the original Penna model, at  $l$ the individual dies, but here the individual has only a $(1-p)$ probability of dying.  The number of individuals who die at age $t$ is just \begin{equation}D(t)=p^Mn(t-M)+\sum_{x=t-M+1}^{t} (1-p)p^{t-x}n(x),
\end{equation} assuming $t\geq M$. The first term ensures that when $t=l+M$, all the remaining people alive with $n(l)$ die, not just a proportion of $1-p$.
\par
It is also possible to combine the original Penna model with the modification described previously, making a combination Penna (CP)  model.  Some people have a genetic tendency for heart disease, or stroke\cite{hassan}. These diseases are not directly affected by senescence.  In a combination of the two models, there would be two classes of death, one instantaneous, as in the original model, and one probability based, as in the proposed modification.  Then each individual is defined in terms of $l$ and $l_P$, where $l$ is the time of a probabilistic mutation and $l_P$ is the time to a Penna mutation, or instant death. $n_{j+1}(0,l,l_P)$ can come from $n_j(x,l',l_P')$ where $l'\geq l$ and $l_P'\geq l_P$.  Note that $l$ cannot be greater than $l_P$.  Therefore, \begin{equation}\begin{aligned}n_{j+1}(0,l,l_P)&=be^{-\alpha l}e^{-\beta l_P}[\sum_x n_j(x,l,l_P)\\
&+(1-e^{-\alpha})(1-e^{-\beta})\sum_{l'>l}\sum_{l_P'>l_P}\sum_x n_j(x,l',l_P')\\
&+(1-e^{-\beta})\sum_{l'>l}\sum_x n_j(x,l',l_P)\\
&+(1-e^{-\alpha})\sum_{l_P'>l_P}\sum_x n_j(x,l,l_P')],\end{aligned}\end{equation} 
where $e^{-\alpha}$ and $e^{-\beta}$ are the probabilities of not getting a senescence or death mutation, respectively.  The death distribution is given by probabilistic and instantaneous deaths, \begin{equation}\begin{aligned}
D(t)=&\sum_{x=t-M}^tp^{t-x}n(x,t)\\&+\sum_{l_P>t}\sum_{l<t}\sum_{x=t-M+1}^{t}p^{t-x}(1-p)n(x,l_P)\\&+p^Mn(t-M,l_P)
\end{aligned}\end{equation}\par
As time passes in the Penna model, the mutations slowly move downwards.  However, there is no evolutionary pressure preventing an individual with $l=0$ from reproducing.  Since there are no positive mutations, my ansatz is that in a steady state, $l=0$ for all individuals.  If $l_P=0$, the individual does not reproduce, so there is an evolutionary pressure for $l_P>0$.  If this happens, then any terms summing over different values of $l$ vanish, so 
\begin{equation}\begin{aligned}
n_{j+1}(0,l,l_P)=&be^{-\alpha l}e^{-\beta l_P}[\sum_{l_P'=l_P}\sum_x n_j(x,l,l_P)\\
&+(1-e^{-\alpha})\sum_{l_P'>l_P}\sum_x n_j(x,l,l_P')].\end{aligned}\end{equation} 
Simplifying in the same manner as in the previous model, the steady state form is the same as eq. 7, except \begin{equation}q_l=p\frac{1-p^{l_P}}{1-p}.\end{equation}  However, since $l=0$, the time until senescence is always 0, eq. 12 is not needed.  The recursive form just gives the age of death.
\section{Results}
\begin{figure}
 \includegraphics[scale=0.35]{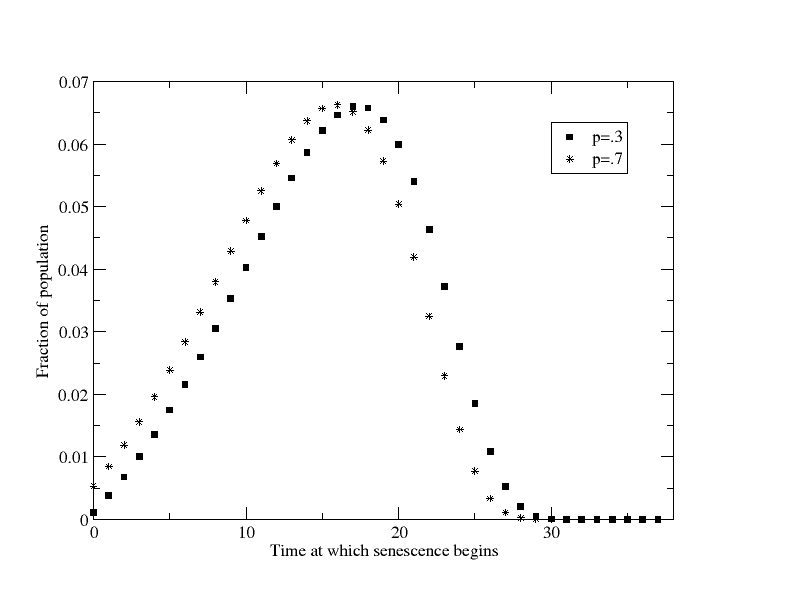}
 \caption{\label{fig1} Higher values of $p$ result in earlier times of senescence in the SP.  In this picture, $M=5$.}
\end{figure}
\begin{figure}
 \includegraphics[scale=0.35]{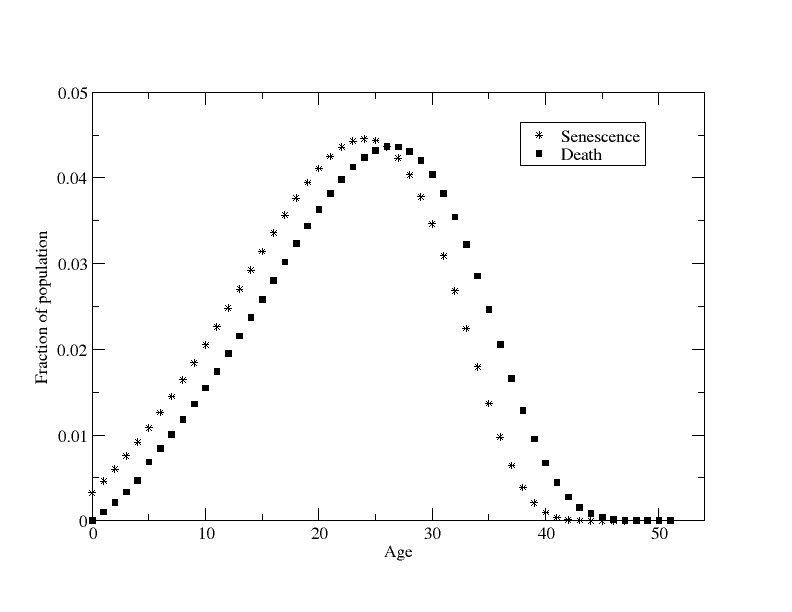}
 \caption{\label{fig2} The death curve in the SP is a shift of the senescence curve, with the amount shifted varying on $p$.  Here $M=5$ and $p=.8$.}
\end{figure}
\begin{figure}
 \includegraphics[scale=0.35]{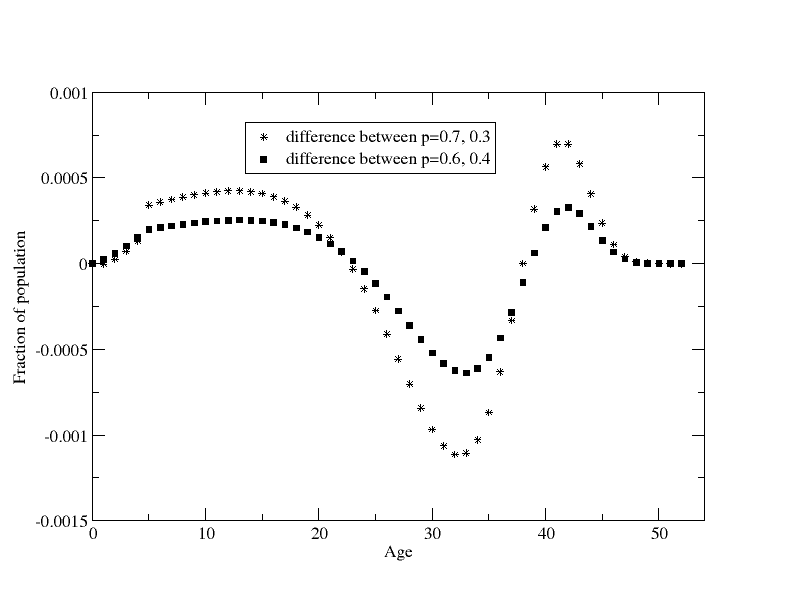}
 \caption{\label{fig3} Higher values of $p$ have a higher proportion of their population reach larger ages, but that proportion difference is countered in the middle of the death distribution. $M=5$}
\end{figure}
Interestingly, in the SP model higher values of $p$ show a ``younger population'' (fig. 2).  With lower $p$ values, an individual with low $l$ will not have as many opportunities to reproduce, since chances are it will die out soon.  This provides an evolutionary pressure for higher values of $l$.  However, if $p$ is high, then individuals with low $l$ can still reproduce.\par
While senescence begins later for populations with lower $p$, death comes earlier once senescence is reached.  The average death time for $n(l)$ is $l+(1-p)+2p(1-p)+3p^2(1-p)+\ldots+(M+1)p^M$, where the last term has no $1-p$ factor, since everybody left alive has to die.  Higher values of $p$ result in later deaths, as would be expected.  The later deaths are more apparent after $l_{max}$, when any living members of the population are undergoing senescence.  Figures 4 and 5 show the differences in percentages, which have to add up to 0.  The increased proportion of older individuals for higher values of $p$ has to be balanced by a reduced proportion of the population at lower ages.  The jump in the differences of death proportions early in figs. 4 and 5 is caused by $M$.  For larger values of $p$, the $1-p$ term in eq. 10 is small enough to make the $p^{t-x}$ term negligible.  However, once $M$ is reached, the final term has no $(1-p)$ factor, causing the jump.\par
The SP model does not show an increase in the mean for higher values of $p$, but it does show that the probability of living to a higher age is greater.  Even though the probability of living to a high age is greater, the maximum age for both populations with higher and lower $p$ is still $l_{max}+M$.  Higher values of $M$ push the time of senescence further forward, since there is less evolutionary incentive for an individual to have a higher $l$.  \begin{figure}
 \includegraphics[scale=0.35]{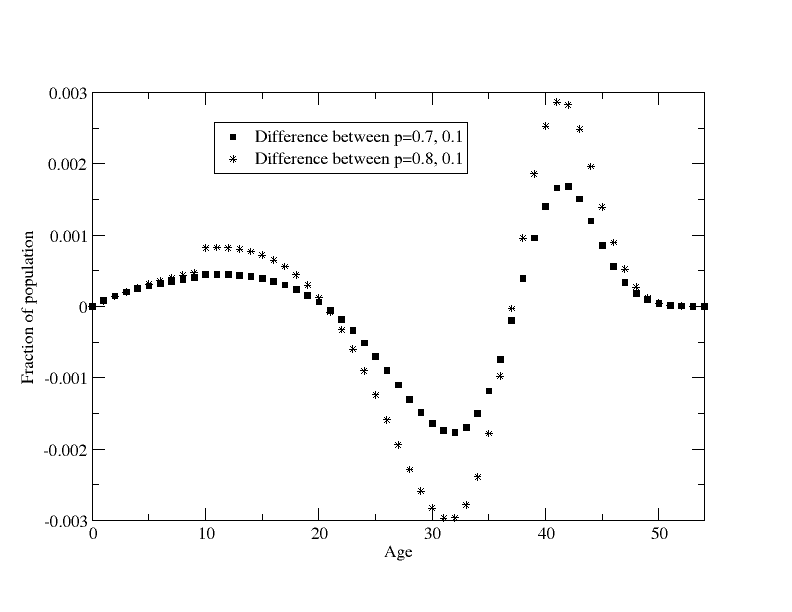}
 \caption{\label{fig4} This shows the same break as fig. 4, but at age 10, since $M=10$ in the SP here.}
\end{figure}
 \begin{figure}
 \includegraphics[scale=0.35]{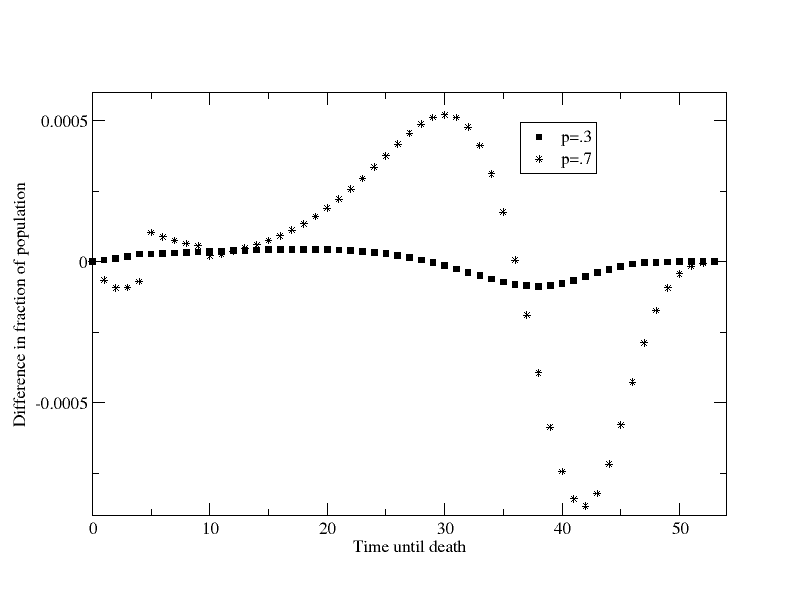}
 \caption{\label{fig5} The differences in the death distribution curves of the SP of $M=5$ and $M=10$, for different values of $p$.}
\end{figure}
\begin{figure}
 \includegraphics[scale=0.35]{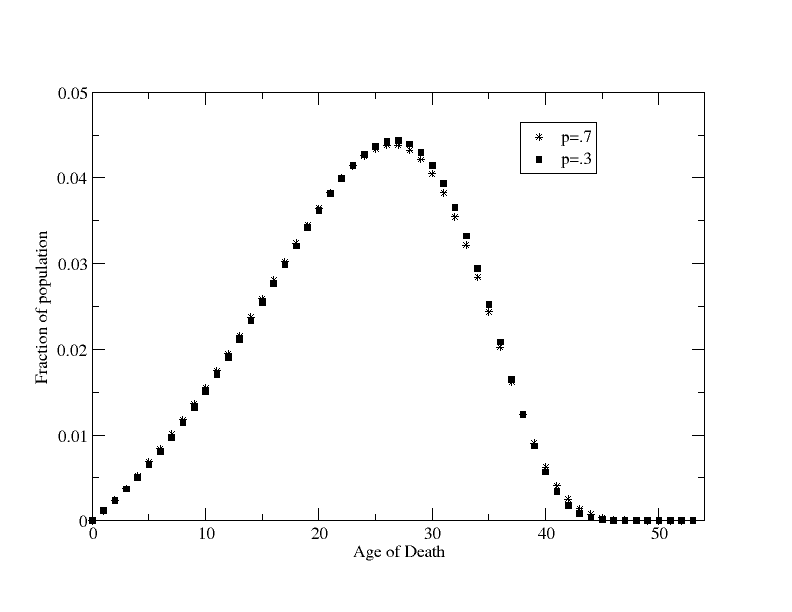}
 \caption{\label{fig6} The death distributions for different values of $p$ look approximately the same, but there is a slightly higher proportion of people alive at later times for higher values of $p$. }
\end{figure}
However, just like $p$, higher values of $M$ also afford a longer time until death, balancing out the earlier senescence times.  Since higher values of $M$ allow for a longer life, the proportion of people alive at a higher age is greater for higher values of $M$.  The breaks in fig. 5 are caused by the same mechanism as the breaks in figs. 3 and 4, except instead of the deaths coming at one specific $M$, they come at the two values of $M$, 5 and 10.\par
\begin{figure}
 \includegraphics[scale=0.35]{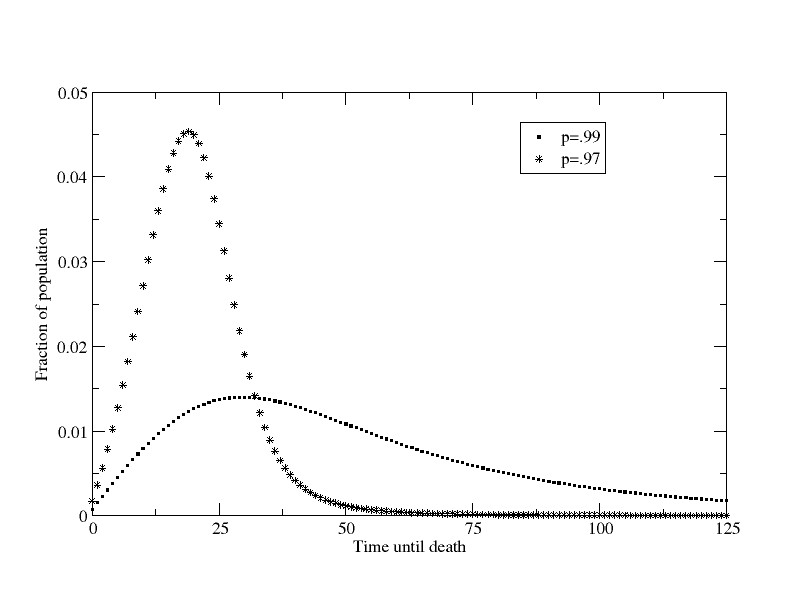}
 \caption{\label{combodiffp}This is the death distribution in the CP, for different values of $p$. Notice that even a slight change in $p$ has a large change on the death distribution. }
\end{figure}
The CP model shows a different structure than the SP model.  It shows a quicker increase in age expectancy, and a slower decrease.  The combination requires a high $p$, because otherwise $q$ degenerates rapidly to $\frac{p}{1-p}$.  If $p$ is too low, then the CP model will not be stable without a Verhulst factor, \begin{equation}V=1-\frac{N(t)}{N_{max}},\end{equation} where $N(t)$ is the number of people alive at time $t$, $N_{max}$ is the maximum allowed number of people, and $V$ is the probability that an individual survives a timestep. This also explains why even small differences in the value of $p$ show large differences in the age distribution curve.
\section{Conclusion}
In this paper, I showed that simple modifications to the Penna model allow for shifts in the lifespan distribution  without changing the maximum lifespan.  Higher values of $p$ in the SP result in younger populations, but they die later.\par
Thinking about the original context for this modification, higher values of $p$ can represent exercising populations.  The maximum age of two populations with different $p$ and the same $M$ is the same, but the exercising population has a higher chance of reaching later ages.  What is interesting, is that this model shows that the exercising population will also be younger.  The times of senescence for high $p$ are lower than low $p$, yet the probability of reaching a high age is greater.  \par
The Penna model is a tool used to help us understand population dynamics.  My modifications of the Penna model take into account senescence of a population, which is a critical part of the aging process, and help to explain the changes in lifespan observed in exercising and non-exercising populations. By adjusting $e^{-\beta}, p,$ and $M$, these should be able to fit actual data.  Further improvements to the models could take into account recent research in autism which suggests that $e^{-\beta}$ is actually a function of time\cite{kong}, and also looking at positive mutations.
\bibliography{bib}

\begin{thebibliography}{10}%
\makeatletter
\providecommand \@ifxundefined [1]{%
 \ifx #1\undefined \expandafter \@firstoftwo
 \else \expandafter \@secondoftwo
\fi
}%
\providecommand \@ifnum [1]{%
 \ifnum #1\expandafter \@firstoftwo
 \else \expandafter \@secondoftwo
\fi
}%
\providecommand \enquote [1]{``#1''}%
\providecommand \bibnamefont  [1]{#1}%
\providecommand \bibfnamefont [1]{#1}%
\providecommand \citenamefont [1]{#1}%
\providecommand\href[0]{\@sanitize\@href}%
\providecommand\@href[1]{\endgroup\@@startlink{#1}\endgroup\@@href}%
\providecommand\@@href[1]{#1\@@endlink}%
\providecommand \@sanitize [0]{\begingroup\catcode`\&12\catcode`\#12\relax}%
\@ifxundefined \pdfoutput {\@firstoftwo}{%
 \@ifnum{\z@=\pdfoutput}{\@firstoftwo}{\@secondoftwo}%
}{%
 \providecommand\@@startlink[1]{\leavevmode\special{html:<a href="#1">}}%
 \providecommand\@@endlink[0]{\special{html:</a>}}%
}{%
 \providecommand\@@startlink[1]{%
  \leavevmode
  \pdfstartlink
   attr{/Border[0 0 1 ]/H/I/C[0 1 1]}%
   user{/Subtype/Link/A<</Type/Action/S/URI/URI(#1)>>}%
  \relax
 }%
 \providecommand\@@endlink[0]{\pdfendlink}%
}%
\providecommand \url  [0]{\begingroup\@sanitize \@url }%
\providecommand \@url [1]{\endgroup\@href {#1}{\urlprefix}}%
\providecommand \urlprefix [0]{URL }%
\providecommand \Eprint[0]{\href }%
\@ifxundefined \urlstyle {%
  \providecommand \doi [1]{doi:\discretionary{}{}{}#1}%
}{%
  \providecommand \doi [0]{doi:\discretionary{}{}{}\begingroup
  \urlstyle{rm}\Url }%
}%
\providecommand \doibase [0]{http://dx.doi.org/}%
\providecommand \Doi[1]{\href{\doibase#1}}%
\providecommand \bibAnnote [3]{%
  \BibitemShut{#1}%
  \begin{quotation}\noindent
    \textsc{Key:}\ #2\\\textsc{Annotation:}\ #3%
  \end{quotation}%
}%
\providecommand \bibAnnoteFile [2]{%
  \IfFileExists{#2}{\bibAnnote {#1} {#2} {\input{#2}}}{}%
}%
\providecommand \typeout [0]{\immediate \write \m@ne }%
\providecommand \selectlanguage [0]{\@gobble}%
\providecommand \bibinfo [0]{\@secondoftwo}%
\providecommand \bibfield [0]{\@secondoftwo}%
\providecommand \translation [1]{[#1]}%
\providecommand \BibitemOpen[0]{}%
\providecommand \bibitemStop [0]{}%
\providecommand \bibitemNoStop [0]{.\EOS\space}%
\providecommand \EOS [0]{\spacefactor3000\relax}%
\providecommand \BibitemShut [1]{\csname bibitem#1\endcsname}%
\bibitem{herbig}%
  \BibitemOpen
  \bibfield{author}{%
  \bibinfo {author} {\bibfnamefont{U.}~\bibnamefont{Herbig}}, \bibinfo {author}
  {\bibfnamefont{W.~A.}\ \bibnamefont{Jobling}}, \bibinfo {author}
  {\bibfnamefont{B.~P.}\ \bibnamefont{Chen}}, \bibinfo {author}
  {\bibfnamefont{D.~J.}\ \bibnamefont{Chen}},\ and\ \bibinfo {author}
  {\bibfnamefont{J.~M.}\ \bibnamefont{Sedivy}},\ }%
  \bibfield{journal}{%
  \bibinfo {journal} {Molecular Cell}\ }%
  \textbf{\bibinfo {volume} {14}},\ \bibinfo {pages} {501} (\bibinfo {month}
  {May}\ \bibinfo {year} {2004})%
  \bibAnnoteFile{NoStop}{herbig}%
\bibitem{degrey}%
  \BibitemOpen
  \bibfield{author}{%
  \bibinfo {author} {\bibfnamefont{A.~D. N.~J.}\ \bibnamefont{de~Grey}},\ }%
  \bibfield{journal}{%
  \bibinfo {journal} {Studies in Ethics, Law, and Technology}\ }%
  \textbf{\bibinfo {volume} {1}},\  (\bibinfo {year} {2007})%
  \bibAnnoteFile{NoStop}{degrey}%
\bibitem{werner}%
  \BibitemOpen
  \bibfield{author}{%
  \bibinfo {author} {\bibfnamefont{C.}~\bibnamefont{Werner}}, \bibinfo {author}
  {\bibfnamefont{T.}~\bibnamefont{Fuerster}}, \bibinfo {author}
  {\bibfnamefont{T.}~\bibnamefont{Widmann}}, \bibinfo {author}
  {\bibfnamefont{J.}~\bibnamefont{Poess}}, \bibinfo {author}
  {\bibfnamefont{C.}~\bibnamefont{Roggia}}, \bibinfo {author}
  {\bibfnamefont{M.}~\bibnamefont{Hanhoun}}, \bibinfo {author}
  {\bibfnamefont{J.}~\bibnamefont{Scharhag}}, \bibinfo {author}
  {\bibfnamefont{N.}~\bibnamefont{Buechner}}, \bibinfo {author}
  {\bibfnamefont{T.}~\bibnamefont{Meyer}}, \bibinfo {author}
  {\bibfnamefont{W.}~\bibnamefont{Kindermann}}, \bibinfo {author}
  {\bibfnamefont{J.}~\bibnamefont{Haendeler}}, \bibinfo {author}
  {\bibfnamefont{M.}~\bibnamefont{Boehm}},\ and\ \bibinfo {author}
  {\bibfnamefont{U.}~\bibnamefont{Laufs}},\ }%
  \bibfield{journal}{%
  \bibinfo {journal} {Circulation}}%
   (\bibinfo {year} {2009})%
  \bibAnnoteFile{NoStop}{werner}%
\bibitem{holloszy}%
  \BibitemOpen
  \bibfield{author}{%
  \bibinfo {author} {\bibfnamefont{J.}~\bibnamefont{Holloszy}},\ }%
  \bibfield{journal}{%
  \bibinfo {journal} {Journal of Applied Physiology}\ }%
  \textbf{\bibinfo {volume} {82}} (\bibinfo {year} {1997})%
  \bibAnnoteFile{NoStop}{holloszy}%
\bibitem{sarna}%
  \BibitemOpen
  \bibfield{author}{%
  \bibinfo {author} {\bibfnamefont{S.}~\bibnamefont{Sarna}}, \bibinfo {author}
  {\bibfnamefont{T.}~\bibnamefont{Sahi}}, \bibinfo {author}
  {\bibfnamefont{M.}~\bibnamefont{Koskenvuo}},\ and\ \bibinfo {author}
  {\bibfnamefont{J.}~\bibnamefont{Kaprio}},\ }%
  \bibfield{journal}{%
  \bibinfo {journal} {Medecine Science Sports and Exercise}\ }%
  \textbf{\bibinfo {volume} {25}} (\bibinfo {year} {1993})%
  \bibAnnoteFile{NoStop}{sarna}%
\bibitem{penna}%
  \BibitemOpen
  \bibfield{author}{%
  \bibinfo {author} {\bibfnamefont{T.}~\bibnamefont{Penna}},\ }%
  \bibfield{journal}{%
  \Doi{10.1007/BF02180147}{\bibinfo {journal} {Journal of Statistical
  Physics}}\ }%
  \textbf{\bibinfo {volume} {78}},\ \bibinfo {pages} {1629} (\bibinfo {year}
  {1995}),\ ISSN \bibinfo {issn} {0022-4715},\
  \url{http://dx.doi.org/10.1007/BF02180147}%
  \bibAnnoteFile{NoStop}{penna}%
\bibitem{stauffer}%
  \BibitemOpen
  \bibfield{author}{%
  \bibinfo {author} {\bibfnamefont{D.}~\bibnamefont{Stauffer}},\ }%
  \bibfield{journal}{%
  \bibinfo {journal} {Bioinformatics and Biology Insights}\ }%
  \textbf{\bibinfo {volume} {1}} (\bibinfo {year} {2007})%
  \bibAnnoteFile{NoStop}{stauffer}%
\bibitem{coe1}%
  \BibitemOpen
  \bibfield{author}{%
  \bibinfo {author} {\bibfnamefont{J.~B.}\ \bibnamefont{Coe}}\ and\ \bibinfo
  {author} {\bibfnamefont{Y.}~\bibnamefont{Mao}},\ }%
  \bibfield{journal}{%
  \Doi{10.1103/PhysRevE.67.061909}{\bibinfo {journal} {Phys. Rev. E}}\ }%
  \textbf{\bibinfo {volume} {67}},\ \bibinfo {pages} {061909} (\bibinfo {month}
  {Jun}\ \bibinfo {year} {2003}),\
  \url{http://link.aps.org/doi/10.1103/PhysRevE.67.061909}%
  \bibAnnoteFile{NoStop}{coe1}%
\bibitem{coe2}%
  \BibitemOpen
  \bibfield{author}{%
  \bibinfo {author} {\bibfnamefont{J.~B.}\ \bibnamefont{Coe}}\ and\ \bibinfo
  {author} {\bibfnamefont{Y.}~\bibnamefont{Mao}},\ }%
  \bibfield{journal}{%
  \Doi{10.1103/PhysRevE.69.041907}{\bibinfo {journal} {Phys. Rev. E}}\ }%
  \textbf{\bibinfo {volume} {69}},\ \bibinfo {pages} {041907} (\bibinfo {month}
  {Apr}\ \bibinfo {year} {2004}),\
  \url{http://link.aps.org/doi/10.1103/PhysRevE.69.041907}%
  \bibAnnoteFile{NoStop}{coe2}%
\bibitem{coe3}%
  \BibitemOpen
  \bibfield{author}{%
  \bibinfo {author} {\bibfnamefont{J.~B.}\ \bibnamefont{Coe}}, \bibinfo
  {author} {\bibfnamefont{Y.}~\bibnamefont{Mao}},\ and\ \bibinfo {author}
  {\bibfnamefont{M.~E.}\ \bibnamefont{Cates}},\ }%
  \bibfield{journal}{%
  \Doi{10.1103/PhysRevLett.89.288103}{\bibinfo {journal} {Phys. Rev. Lett.}}\
  }%
  \textbf{\bibinfo {volume} {89}},\ \bibinfo {pages} {288103} (\bibinfo {month}
  {Dec}\ \bibinfo {year} {2002}),\
  \url{http://link.aps.org/doi/10.1103/PhysRevLett.89.288103}%
  \bibAnnoteFile{NoStop}{coe3}%
\bibitem{hassan}%
  \BibitemOpen
  \bibfield{author}{%
  \bibinfo {author} {\bibfnamefont{A.}~\bibnamefont{Hassan}}\ and\ \bibinfo
  {author} {\bibfnamefont{H.~S.}\ \bibnamefont{Markus}},\ }%
  \bibfield{journal}{%
  \Doi{10.1093/brain/123.9.1784}{\bibinfo {journal} {Brain}}\ }%
  \textbf{\bibinfo {volume} {123}},\ \bibinfo {pages} {1784} (\bibinfo {year}
  {2000}),\
  \Eprint{http://arxiv.org/abs/http://brain.oxfordjournals.org/content/123/9/1%
784.full.pdf+html}{http://brain.oxfordjournals.org/content/123/9/1784.full.pdf%
+html},\ \url{http://brain.oxfordjournals.org/content/123/9/1784.abstract}%
  \bibAnnoteFile{NoStop}{hassan}%
\bibitem{kong}%
  \BibitemOpen
  \bibfield{author}{%
  \bibinfo {author} {\bibfnamefont{A.}~\bibnamefont{Kong}}, \bibinfo {author}
  {\bibfnamefont{M.~L.}\ \bibnamefont{Friggeand}}, \bibinfo {author}
  {\bibfnamefont{G.}~\bibnamefont{Masson}}, \bibinfo {author}
  {\bibfnamefont{S.}~\bibnamefont{Besenbacher}}, \bibinfo {author}
  {\bibfnamefont{P.}~\bibnamefont{Sulem}}, \bibinfo {author}
  {\bibfnamefont{G.}~\bibnamefont{Magnusson}}, \bibinfo {author}
  {\bibfnamefont{S.~A.}\ \bibnamefont{Gudjonsson}}, \bibinfo {author}
  {\bibfnamefont{A.}~\bibnamefont{Sigurdsson}}, \bibinfo {author}
  {\bibfnamefont{A.}~\bibnamefont{Jonasdottir}}, \bibinfo {author}
  {\bibfnamefont{A.}~\bibnamefont{Jonasdottir}}, \bibinfo {author}
  {\bibfnamefont{W.~S.~W.}\ \bibnamefont{Wong}}, \bibinfo {author}
  {\bibfnamefont{G.}~\bibnamefont{Sigurdsson}}, \bibinfo {author}
  {\bibfnamefont{G.~B.}\ \bibnamefont{Walters}}, \bibinfo {author}
  {\bibfnamefont{S.}~\bibnamefont{Steinberg}}, \bibinfo {author}
  {\bibfnamefont{H.}~\bibnamefont{Helgason}}, \bibinfo {author}
  {\bibfnamefont{G.}~\bibnamefont{Thorleifsson}}, \bibinfo {author}
  {\bibfnamefont{D.~F.}\ \bibnamefont{Gudbjartsson}}, \bibinfo {author}
  {\bibfnamefont{A.}~\bibnamefont{Helgason}}, \bibinfo {author}
  {\bibfnamefont{O.~T.}\ \bibnamefont{Magnusson}}, \bibinfo {author}
  {\bibfnamefont{U.}~\bibnamefont{Thorsteinsdottir}},\ and\ \bibinfo {author}
  {\bibfnamefont{K.}~\bibnamefont{Stefansson}},\ }%
  \bibfield{journal}{%
  \bibinfo {journal} {Nature}\ }%
  \textbf{\bibinfo {volume} {488}} (\bibinfo {month} {Aug}\ \bibinfo {year}
  {2012}),\ \doi{\bibinfo {doi} {10.1038/nature11396}}%
  \bibAnnoteFile{NoStop}{kong}%
\end{thebibliography}%
\end{document}